\documentclass[conference]{IEEEtran}
\usepackage{graphicx}
\usepackage{amsmath}
\usepackage{amssymb}

\begin{document}
\title{Almost-Optimum Signature Matrices in Binary-Input Synchronous Overloaded CDMA}

\author{M.~Heidari~Khoozani,~A.~Rashidinejad,~M.~H.~Lotfi~Froushani,~P.~Pad,~F.~Marvasti\\
Advanced Communication Research Institute, Department of Electrical Engineering\\
Sharif University of Technology, Tehran, Iran\\
\{heydari,arashidinejad,lotfi,pedram\_pad\}@ee.sharif.edu \& marvasti@sharif.edu} 

\maketitle

\IEEEpeerreviewmaketitle

\begin{abstract}
The everlasting bandwidth limitations in wireless communication networks has directed the researchers' thrust toward analyzing the prospect of overloaded Code Division Multiple Access (CDMA). In this paper, we have proposed a Genetic Algorithm in search of optimum signature matrices for binary-input synchronous CDMA. The main measure of optimality considered in this paper, is the per-user channel capacity of the overall multiple access system. Our resulting matrices differ from the renowned Welch Bound Equality (WBE) codes, regarding the fact that our attention is specifically aimed at binary, rather than Gaussian, input distributions. Since design based on channel capacity is computationally expensive, we have focused on introducing a set of alternative criteria that not only speed up the matrix formation procedure, but also maintain optimality. The Bit Error Rate (BER) and Constellation measures are our main criteria propositions. Simulation results also verify our analytical justifications.
\end{abstract}

\section{Introduction}\label{sec:intro}

Since the dawn of synchronous Code Division Multiple Access (CDMA), there has been an everincreasing demand to improve their superior capabilities, such as channel capacity, Quality of Service (QoS) and the per-user data bit rate. One main obstacle in the path of system engineers, however, is to increase the number of CDMA channel users while sustaining reliable transmission. Of course, if the number of users is less than or equal to the available spreading factor, the utilization of completely orthogonal codes, e.g. Hadamard codes, leads to satisfaction in the matter. Nevertheless, no such orthogonal codes are attainable for an overloaded CDMA system; thus errors caused due to Multi Access Interference (MAI) as well as inherent system noise are inevitable.

The analysis of the overloaded CDMA concept comes to life when channel bandwidth limitations are realized, therefore a great deal of research has been aimed at the subject in recent years. For example \cite{Verdu}-\cite{Grant} have focused on pseudonoise (PN) spreading, \cite{Sari} has presented PN/OCDMA (PN/O) signature sets and \cite{PedramIT}-\cite{PedramISIT} have presented well-behaved signature sets for binary-input\footnote{By binary-input, we mean that the input vector to the system is assumed to be binary, i.e. $\pm 1$'s.} CDMA channels, namely COW matrices and \cite{KasraIT}-\cite{Enigma} have evaluated the overloaded noiseless and noisy channel capacity for different orders of dimension.

The renowned Welch Bound Equality (WBE) codes have been introduced in \cite{Welch} and justified to reach maximum capacity for Gaussian input distributions \cite{WBE1}-\cite{WBE2}. However, regarding the fact that this paper's attention is specifically on binary input signals, the capacity-reaching justifications for WBE's are no longer true. As a matter of fact, according to simulation results, WBE's exhibit diverse performances for binary inputs and in most cases below optimum characteristics.

Among the literature, \cite{PedramIT}-\cite{PedramISIT} are the only that deal with binary-input overloaded CDMA. These papers have proposed the construction algorithms for matrices with binary ($-1$ and $1$), ternary ($-1$, $0$ and $1$) and generally limited elements in the CDMA signature sequences. Although limiting the elements of the signature matrix causes design simplicity, it usually leads to optimality limitations. Conversely, in this paper, we have not confined the elements of the optimum signature matrices, apart from the fact that they should be real.

It is noteworthy that none of the binary-input signatures schemes that have been
proposed in the literature (including the WBE and COW matrices) guarantee
optimality regarding to channel capacity maximization in different SNRs\footnote{Most of the work done in the literature assume noiseless and high SNR conditions rather than also evaluating their propositions in relatively low SNRs.}.

In this paper, we basically pursued the following objectives:
\begin{itemize}
\item Evaluating the key equations relating any signature sequence matrix with its corresponding binary-input synchronous CDMA channel capacity.
\item Introducing a set of appropriate alternative criteria, instead of the channel capacity itself, that could be optimized over signature matrices in order to simplify and speed up the procedure of attaining optimum capacity code sets.
\item Optimizing the proposed criteria using a real-valued Genetic Algorithm and comparing the results with analytic evaluations.
\end{itemize}


In section \ref{sec:cap} we have presented the basic preliminaries of our research. The general definition of a binary-input AWGN CDMA channel, followed by an outline of our main concern, i.e., the channel capacity maximization, is presented. Hence, with the results of the formulations presented in this section, the precise channel capacity could be obtained for any specific signature matrix. 

In section \ref{sec:criteria}, some criteria for the optimality of a signature code matrix are presented. These measures are proved to be in coherence with optimum (or in some simplifications sub-optimum) capacity characteristics and prepare the grounds for rapid selection of the signature matrix set for the overloaded system. The main setback in choosing the perfect measure is that a compromise must be solved between each criterion's complexity and its optimality performance. Nevertheless, our proposed criteria not only maintain near-optimum characteristics, but decrease the computation order to a great extent.

There are two main approaches that have been proposed in maximizing the overloaded sum channel capacity, the direct channel capacity maximization and the indirect Bit Error Rate (BER) minimization schemes. The indirect scheme is based on our conjecture that matrices that exhibit low BER characteristics, also behave near-optimally regarding to channel capacity maximization. Based on the second set of criteria, in subsection \ref{subsec:constellation} we have justified the proposition of a set of Constellation measures\footnote{Constellation measures deal with the possible output projection points in the image field for noiseless conditions.}.

In section \ref{sec:compare} we have carried out the Genetic optimization Algorithm in order to execute and deeply compare all the priorly explained criteria optimizations. Finally the conclusions and future works are depicted in section \ref{sec:conclusion}.

\section{Problem Definition and Preliminaries}\label{sec:cap} 
Synchronous CDMA systems, regardless of the fact that they are overloaded or not, are formulated in a consistent manner as follows,
\begin{equation}\label{equ:model}
Y={\bf A}X+N
\end{equation}
where ${\bf A}$ is an $m \times n$ matrix with CDMA signature columns. Due to the fact that this paper's concentration is basically on overloaded synchronous CDMA systems, the number of users ($n$) is larger than the signature spreading factor ($m$). Also, $X$ is the user column vector with entries $\{\pm 1\}$ and $N$ is the AWGN vector $N=[N_1,\ldots ,N_m]^T$, such that $N_i$'s are i.i.d random variables.

We shall also denote the overloading factor of such a system as,

\begin{equation}
\beta = \frac{n}{m}
\end{equation}

In general, the elements of ${\bf {\bf A}}$ are free to take any real or complex value, but in our focus, ${\bf A}$ should be composed of real-valued elements with the only restriction of having normalized signature columns to promote fairness between all users.

It is clear that the sum channel capacity for a given $\bf A$ matrix is as below,
\begin{equation}\label{equ:CAA}
C({\bf A})=\max_{p(X)}{I(X;Y)}
\end{equation}
Note here, that the maximization is over all possible multiplicative distributions on $X$. However, according to the conjectures mentioned in \cite{PedramIT} and \cite{KasraIT}, for every matrix $\bf A$, the capacity is met for a uniform input distribution. Thus we have,
\begin{equation}\label{equ:CA}
C({\bf A})=I(X;Y)~,~p(X=\bar{X})=\frac{1}{2^n}
\end{equation}
In order to facilitate our path towards generating the optimum overloaded CDMA matrices, we need to initially evaluate the  maximum channel capacity. We define the maximum channel capacity as,
\begin{equation}\label{equ:cap}
C(n,m,\sigma_N)=\max_{{\bf A}\in \mathbf{R}_{m\times n}}{C({\bf A})}
\end{equation}
Taking into account that ${\bf A}$ is deterministic, we have,
\begin{equation}\label{equ:info}
I(X;Y)=h(Y)-h(N)
\end{equation}
Also, keeping in mind that the Gaussian noise elements are independent, $f_N(N)$ is as follows,
\begin{eqnarray}\label{equ:fn}
f_N(N)=\left(\frac{1}{2\pi \sigma_N^2 }\right)^{\frac{m}{2}}\times\prod_{i=1}^m{\exp\left( \frac{N_i^2}{2\sigma_N^2}\right)}
\end{eqnarray}
\newtheorem{mytheorem}{\textbf{Theorem}}
Based on \cite{KasraIT} and due to the uniform distribution of $X$, the probability distribution of $Y$ can be computed as below,
\begin{align}\label{equ:fy}
f_{Y}&(Y ) =\frac{1}{2^n} \times\nonumber\\
&\sum_{\bar{X}\in \{\pm 1\}^{n\times 1}}{\Bigg[\Big(\frac{1}{2\pi \sigma_N^2 }\Big)^{\frac{m}{2}}\prod_{i=1}^m \exp\left(\frac{\left(y_i-{\bf A}_{i} \cdotp \bar{X}\right)^2}{2\sigma_N^2}\right)\Bigg]}
\end{align}

%
%


where $y_i$ and ${\bf A}_{i}$ are the $i^{th}$ entry of $Y$ and the $i^{th}$ row of matrix ${\bf A}$, respectively. In addition, the summation is over all possible $\bar{X}\in \{\pm 1\}^{n\times 1}$.
%
%
%
%
%
%
%
%

According to (\ref{equ:fy}), we can have $h_Y(Y)$,
\begin{align}\label{equ:h_y}
h_Y(Y)=\underbrace{\int \ldots \int}_m f_Y(Y)\log_2(f_Y(Y ))\mathrm{d}y_1\ldots\mathrm{d}y_m
\end{align}
Note that in order to evaluate $C({\bf A})$, we need to subtract $h(N)$ from (\ref{equ:h_y}), where $h(N)$ can also be computed using the definition of differential entropies.

It's noteworthy to mention that using the relations mentioned above, one can evaluate the sum capacity of any arbitrary matrix, which is the basis of our comparisons throughout the paper. According to this achievement, we have the ability to compare various proposed matrices regarding to the optimality of their channel capacity.

In the next section, we have proposed the different possible criteria and methods in choosing optimum signature matrices for different dimensions and SNRs. The schemes are presented and deeply evaluated under different scenarios, such as the Capacity, BER and Constellation Criteria.

\section{Optimum Matrices Criteria}\label{sec:criteria}
As priorly explained, we have tried to match the optimality of a signature matrix capacity with certain criteria, so that it would be possible to find an optimum matrix with arbitrary dimensions and SNR. This approach helps the system designer to easily implement an overloaded CDMA structure without the fuss of probing blindly into any feasible matrix.
To put it straight, in this approach the system designer just has to optimize the certain criterion, while being sure of the sub-optimality (if not optimality) of the generated signature matrix.
\subsection{The Capacity Scenario}\label{subsec:cap}
Now that we have provided the grounds for evaluating the sum channel capacity of an overloaded synchronous CDMA system, we propose that the most reliable measurement criterion is the channel capacity itself.
However, due to the fact that pursuing this basis requires excessive complexity in both calculations and time, we have just relied on the results of this section as a comparison measure throughout the paper.

\begin{figure}[t]
\centering
\includegraphics[width=9cm]{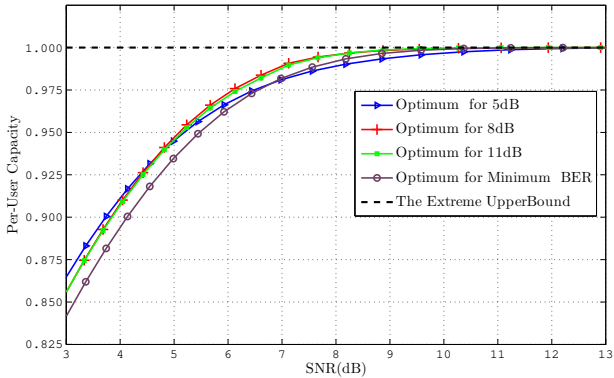}
\caption{The Per-User Channel Capacity of Optimum Code Matrices (Capacity and BER) with $\beta = \frac{4}{3}$ Versus Different $SNR$s}
\label{fig:cap}
\end{figure}

Fig. \ref{fig:cap} illustrates the condition of $3$ optimum matrices for an overloading factor of $\frac{4}{3}$, each optimized for a certain SNR, in an SNR sweep observation. As is clear, the matrices exhibit a channel capacity close to the extreme upper bound, i.e. $1$, even in low SNRs. It is noteworthy to mention that the per-user channel capacity is defined as the ratio of the sum channel capacity to the number of users, keeping in mind that the channel is fairly divided among all users.

The calculation cost of the capacity criterion is so immense that it cannot be extended to wide dimension ranges, thus the need for more simpler criteria comes to life. As previously explained, this is the main focus of this research. Henceforth, we have basically pursued a stagewise procedure in search of these alternate criteria, which immensely contribute to the possibility of designing highly reliable CDMA systems. 

In the next subsection, we have provided the Bit Error Rate criterion as a better-handled criterion than that of capacity. The BER, not only enables extensive search dimensions, but is observed to be reversely related to CDMA channel capacity.

\subsection{The BER Scenario}\label{subsec:BER}
In this subsection we have introduced another measure to find the optimum matrix, i.e. the Bit Error Rate. The optimality of a CDMA signature matrix could be based on both maximum channel capacity or minimum bit error rate. In our research, we are only concerned with the prior of the two standards, however, it seems that they both are closely bonded in a reverse relationship. By this, we conjecture that the lower the BER a certain signature matrix demonstrates, the higher its channel capacity will be. Using Genetic Algorithm to minimize the BER leads to an optimum matrix. Nevertheless, in order to implement a BER evaluation, we will have to utilize a channel decoder. In fact, we have opted the ML\footnote{Maximum Likelihood} decoder throughout this research.


According to the aforementioned discussions, we carried out the same genetically based algorithm in order to determine optimum matrices with respect to minimum BER. 



It is apparent from Fig. \ref{fig:cap} that our conjectures are in consent with reality. The minimum BER matrix adheres to the approximate upper bound for capacity, found in the previous section. This observation clarifies the fact that in order to be able to find higher-dimensioned code matrices, we could exploit the BER as a better measure than the exhaustive capacity computations.

Although calculating the BER involves much less complexity than the capacity itself, it still lacks the feasibility to be implemented on practical orders of dimension. It is true that in order for the BER to be reversely correlated to the capacity, a large array of bits must be produced, coded, sent through the channel and decoded at the receiver. The lower the number of evaluated bits, the more dissimilar BER and channel capacity tend to be. To be more precise, a single bit transmitted through an AWGN channel, needs to be compared with all the possible output constellation points to ensure ML detection.

In the next section a faster and more practical set of standards have been presented, that slightly adjust the trade-off between optimality and implementability in favor of the latter.

\subsection{Output Constellation Schemes}\label{subsec:constellation}

The abovementioned calculations induces the motive to propose some less complex, but optimum criterion which could be easily implemented; hence we altered our viewpoint towards the constellation of noiseless output vectors in the image field. 


In this subsection, we focus on maximizing the capacity with respect to the constellation of noiseless output vectors. Throughout this paper we consider the output constellation points as,
\begin{equation}
Z_{i}=AX_{i}\quad X_1,X_2,\cdots,X_{2^n}\in\{ \pm 1 \}^{n\times 1}
\end{equation} 


At first, we focus on the minimum distance criterion, which is widely accepted among coding theory researchers. In this scheme, we have proposed that if we want to obtain an optimum matrix regarding to the capacity, we must maximize the minimum distance between the noiseless $Y$'s.  Interestingly enough, the same measure, i.e. the minimum distance maximization, yields sub-optimum results in our point of concern.

The following relation explicitly explains this criterion,
\begin{equation}\label{equ:mindis}
\nu_1\ =\min_{i \neq j}\ ||Z_i-Z_j||
\end{equation}
where $||U||$ represents the Euclidean norm of the vector $U$. We shall denote $\nu_1$ as the \textit{Minimum Distance} (MD) criterion throughout this text. Note that $\nu_1$ must be maximized in order to find a sub-optimum matrix. 
\begin{figure}[t]
\centering
\includegraphics[width=9cm]{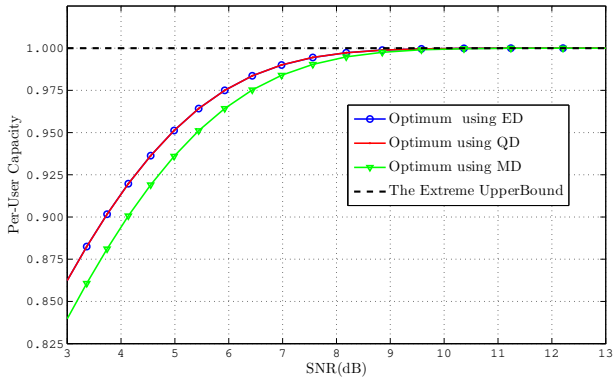}
\caption{Comparison Between the Sum Channel Capacity of the Different Constellation Methods Matrices ($\nu_1$-$\nu_3$) Versus Different $SNR$'s ($\beta = \frac{4}{3}$)}
\label{fig:GA_DIS}
\end{figure}

To take a more analytical approach, the upper bound for error probability is needed. Therefore the independency of $x_i$'s results in,
\begin{equation}
P_e=\frac{1}{2^n}\sum_{i=1}^{2^n}{P\bigg{(}\bigcup_{\begin{smallmatrix} k=1 \\ k\neq i \end{smallmatrix}}^{2^n}||Y_i-Z_i||^2 > ||Y_i- Z_k||^2\bigg{)}}
\end{equation}
Since the noise vector elements are \textit{i.i.d}  Gaussian random variable with variance $\sigma_N$, the error probability upper bounded as,
\begin{equation}\label{equ:se7en}
P_e \leq \frac{1}{2^n}\sum_{i=1}^{2^n}{\sum_{\begin{smallmatrix} j=1 \\ j\neq i \end{smallmatrix}}^{2^n}Q\left(\frac{||Z_i-Z_j||}{2\sigma_N}\right)}
\end{equation}
where $Q(x)$ is the Q-function, the Tail Probability of a zero mean Gaussian random variable with variance $1$.  

According to (\ref{equ:se7en}), the \textit{Q Distance} (QD) measure is introduced as follows,
\begin{equation}
\nu_2\ =\ \sum_{i=1}^{2^n}{\sum_{\begin{smallmatrix} j=1 \\ j\neq i \end{smallmatrix}}^{2^n}Q\left(\frac{||Z_i-Z_j||}{2\sigma}\right)}
\end{equation}
It must be considered that $\nu_2$, as the QD criterion, should be minimized in order to find optimum code matrices.

\begin{figure}[t]
\centering
\includegraphics[width=9cm]{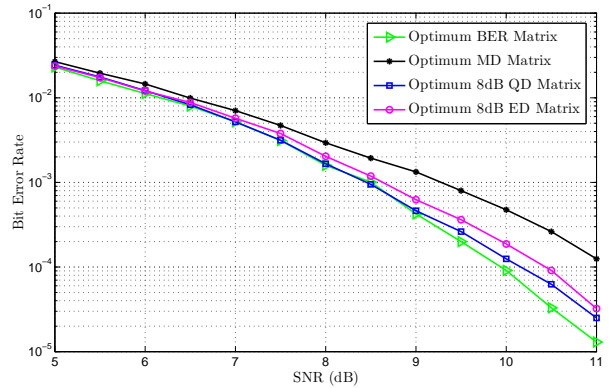}
\caption{Comparison Between the Bit Error Rate of the Different Constellation Methods Matrices ($\nu_1$, $\nu_2$ and $\nu_3$) and the Optimum BER Matrix Versus Different $SNR$'s ($\beta = \frac{4}{3}$)}
\label{fig:BER_SWEEP}
\end{figure}

The \textit{Q}-function that appears in $\nu_2$ can be approximated with the less complex equation as below,
\begin{equation}\label{equ:curvfit}
Q(x)\ \simeq\ a \exp{\left(-\left(\frac{x+b}{c}\right)^2\right)}
\end{equation}
where curve fitting results in $a=0.7$, $b=1$ and $c=1.6$.

According to (\ref{equ:curvfit}), $\nu_2$ can be simplified as,
\begin{equation}\label{equ:ExpDist}
\nu_3\ =\ \sum_{i=1}^{2^n}{\sum_{\begin{smallmatrix} j=1 \\ j\neq i \end{smallmatrix}}^{2^n}\exp\Bigg[-\left(\frac{\frac{||Z_i-Z_j||}{2\sigma_N}+1}{1.6}\right)^2\Bigg]}
\end{equation}
where $\nu_3$ refers to as the \textit{Exponential Distance} (ED) criterion, which must be minimized.


In high SNRs equation (\ref{equ:ExpDist}) can be reduced to only one exponential element which corresponds to the Minimum Distance; consequently we expect the similarity between MD and ED in high SNRs followed by the advantage of ED in low SNRs.     

In Figs. \ref{fig:GA_DIS} and \ref{fig:BER_SWEEP} the mentioned criteria, i.e. MD, QD, and ED, optimum matrices are compared together in  separate capacity and BER plots. As can be seen, ED and QD have approximately the same performance, which demonstrates the perfect approximation in (\ref{equ:curvfit}). This provides the system designer with an advantage of less computational complexity in choosing perfect signature matrices alongside negligible behavioral offshoot. In addition, these two measures perform quite better than the simpler MD method, which can be sheerly proved due to the excessive complexity of the prior two schemes.

In the next section, we have illustrated the optimality characteristics of the different proposed criteria, which is followed by in-depth comparisons between these schemes. The trade-off between matrix optimality and computational complexity is perfectly clear from analytical as well as simulation results.

\section{Comparison and Results}\label{sec:compare}

\begin{figure}[t]
\centering
\includegraphics[width=9cm]{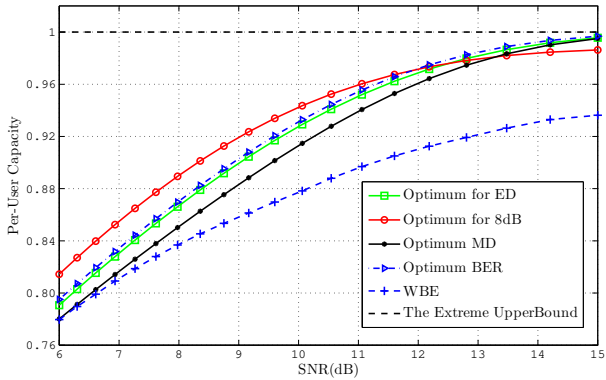}
\caption{Comparison Between the Sum Channel Capacity of the Different Criteria Matrices Versus Different $SNR$'s ($\beta = \frac{5}{2}$)}
\label{fig:comparison}
\end{figure}

As an aftermath to the achieved criteria in the previous sections, we carried out their optimization via Genetic Algorithms. It is obvious that for each $n$, $m$ and SNR, we expect a particular, and probably inconsistent, signature matrix to fulfill the optimality criterion. However, it was observed, and then certified, that optimum matrices in the same dimensions usually exhibit near-optimum, if not optimum, characteristics in other SNRs, hence providing perfect signature matrices for a wide extent of noise powers. Fig. \ref{fig:comparison} contains the per-user capacity versus SNR for the matrices optimized with capacity, BER, MD and ED criteria compared with a Welch Bound Equality (WBE) matrix set. As it is anticipated, because of optimality of WBE matrices for Gaussian-input distribution, there is no guarantee to be optimum as far as we are concerned. Note that the overloading limitation sets the extra restriction that such systems will fall short of the maximum feasible per-user capacity, that is $1$, for low SNRs. 

It should be noted that in order to be able to compare the results of our criteria with the achievable capacity, we were obliged to evaluate the results in low orders of dimensions. However, due to the satisfying characteristics of the matrices found by optimizing these criteria, it can be predicted that optimizing these criteria in larger dimensions also provides near-optimum matrices.
 
It is worth to mention that we have disregard the QD measure, due to exactly similar results to the simpler ED method. Considering Fig.\ref{fig:comparison}, it is clear that the optimum matrix for $8$ dB is not optimum for other SNRs, especially high SNRs, so its falling below other matrices is not an issue.  

It can also be observed that the BER and ED criteria have relatively near optimum performance, even though the implementation is obtained with less complexity. Considering the fact that these two methods minimize the Bit Error Rate, i.e. error probability, the conjecture proposed in section \ref{sec:criteria} is confirmed. Remarkably enough, despite the fact that ED and BER extremely differ in computational complexity, i.e. estimating the BER is more challenging due to the simulation of the channel and randomness behavior; ED and BER demonstrate the same performance regarding to sum channel capacity.

MD, which is proved to be a specialization of the general ED, thusly falls beneath the other criteria in the figure. This sub-optimality may be thought to be compensated with the simplicity of MD. Of course, as anticipated, in high SNRs the MD and ED criteria tend to converge; providing an extra degree of freedom for the overloaded system. 

\begin{figure}[t]
\centering
\includegraphics[width=9cm]{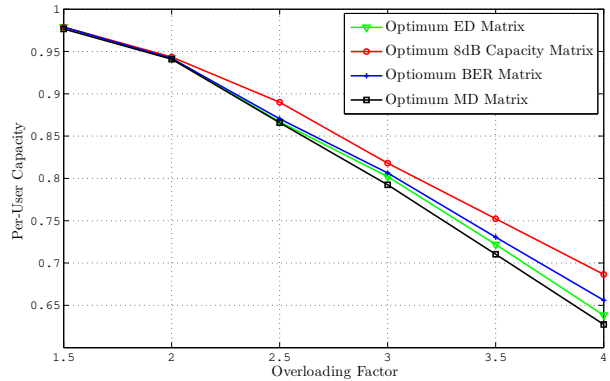}
\caption{Per-user Capacity of Different Proposed Criteria Versus a Range of Overloading Factors. (In SNR 8dB)}
\label{fig:overload}
\end{figure}

Another important feature of our proposed criteria-based technique is that it can lead to results with an extremely flexible behavior in different overloading factors ($\frac{n}{m}$), thus upheaving the implementability of overloaded CDMA systems. Therefore, we have presented the results of the different criteria's per-user capacity while changing the overloading factor in Fig. \ref{fig:overload}. Such a feature brings about the extension of our approach's usage in limited available chip-rates, i.e. small $m$. It can be seen, the fact that we did not restrict our signature matrix elements in all the different scenarios, leads to us enabling high capacity quantities while the overloading factors are relatively high. Although the per-user capacity still decreases as the overloading factor grows, it descends more smoothly in our case; sustaining an acceptable range for larger overloads.

\section{Conclusion and Future Works}\label{sec:conclusion}

In this paper, we proposed a method to determine the capacity related to a specific matrix, then in order to find optimum capacity matrices, a number of criteria have been proposed. These measures differ in optimality and complexity with one another. Generally, the proposed Exponent Distance (ED) criterion presents the best performance regarding the complexity and optimality. It should be stated that to find the optimum matrices in each criteria, we implemented a novel Genetic Algorithm approach. The real valued matrices found by this method discriminate our approach with the prior researches, providing outstanding capability in specific cases in which the chip rate is limited. Considering this feature, the optimum matrices tend to sustain an appropriate per-user capacities for high overloading factors.

The algorithmic procedure in providing the optimum overloaded signature matrices as well as precise analytical evaluations are appropriate future works in the aforementioned research area. Such algorithms may be utilized to gain even more time efficiency while designing an optimum overloaded code division multiple access system. In addition, implementing a similar approach for Optical CDMA systems in presences of Poisson noise is considerable.  

\bibliographystyle{IEEEtran}
\bibliography{paper_ieee}

\end{document}